\documentclass[%
 reprint,
 amsmath,amssymb,
 aps,
 prl,
floatfix,
]{revtex4-2}

\usepackage[mode=buildnew]{standalone}
\usepackage{graphicx}
\usepackage{dcolumn}
\usepackage{bm}
\usepackage{braket}
\usepackage{xcolor}
\usepackage[utf8]{inputenc}
\usepackage[T1]{fontenc}
\usepackage[english]{babel}
\usepackage{letltxmacro}

\LetLtxMacro{\ORIGselectlanguage}{\selectlanguage}
\makeatletter
\DeclareRobustCommand{\selectlanguage}[1]{%
  \@ifundefined{alias@\string#1}
    {\ORIGselectlanguage{#1}}
    {\begingroup\edef\x{\endgroup
       \noexpand\ORIGselectlanguage{\@nameuse{alias@#1}}}\x}%
}
\newcommand{\definelanguagealias}[2]{%
  \@namedef{alias@#1}{#2}%
}

\definelanguagealias{en}{english}

\begin{document}

\preprint{APS/123-QED}

\title{Time-resolved statistics of snippets as general framework\\ for model-free entropy estimators}% Force line breaks with \\

\author{Jann van der Meer}
\thanks{J.v.d.M. and J.D. contributed equally to this work.}
\author{Julius Degünther}
\thanks{J.v.d.M. and J.D. contributed equally to this work.}
\author{Udo Seifert}
\affiliation{
 II. Institut für Theoretische Physik, Universität Stuttgart, 70550 Stuttgart, Germany
}

\date{\today}

\begin{abstract}
Irreversibility is commonly quantified by entropy production. An external observer can estimate it through measuring an observable that is antisymmetric under time-reversal like a current. We introduce a general framework that, {\sl{inter alia}}, allows us to infer a lower bound on entropy production through measuring the time-resolved statistics of events with any symmetry under time-reversal, in particular, time-symmetric instantaneous events. We emphasize Markovianity as a property of certain events rather than of the full system and introduce an operationally accessible criterion for this weakened Markov property. Conceptually, the approach is based on snippets as particular sections of trajectories, for which a generalized detailed balance relation is discussed.
\end{abstract}

\maketitle

\begin{figure*}[t]
\centering
%\includestandalone[scale=1]{Tikz_symmlink_v3.tex}
\includegraphics[scale=1]{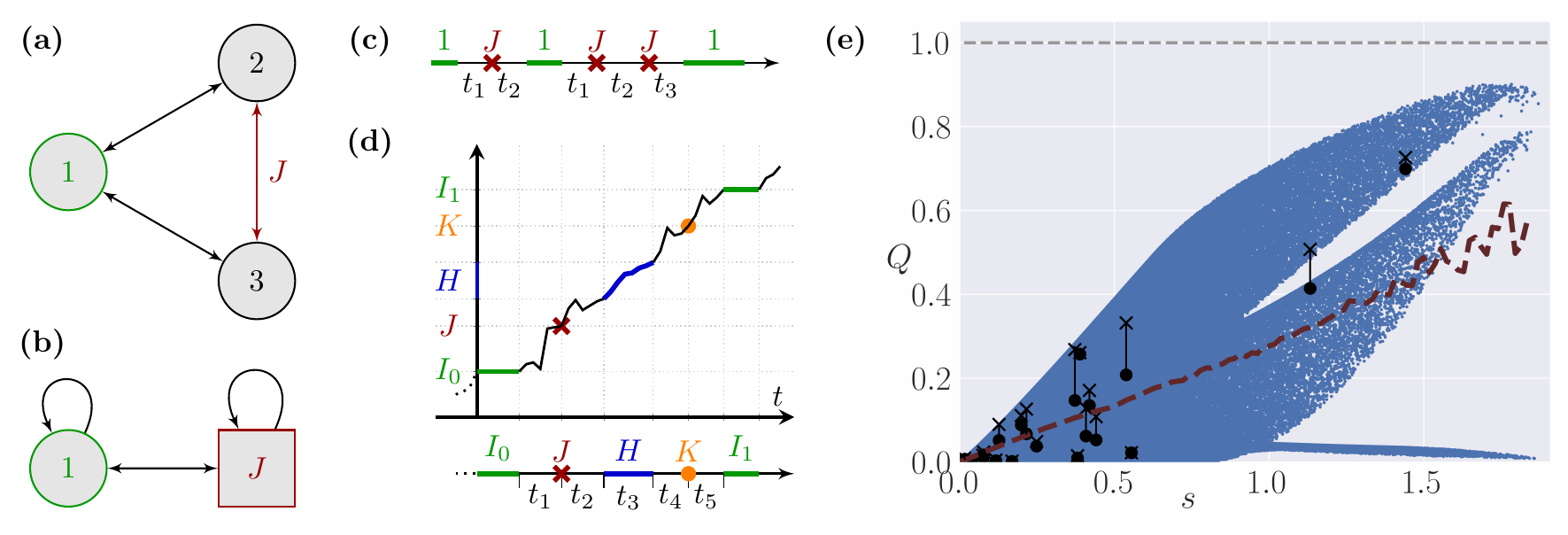}
\caption[cg1]{Entropy estimation based on time-resolved statistics. a) Paradigmatic three-state Markov network. We assume that only state $1$ and transitions along the edge $J$, but not their directions, can be observed. b) Effective network of the coarse-grained description in which $J$ is not a state. c) Time series of observed events. It shows two possible trajectory snippets with corresponding waiting times. d) A trajectory snippet from a network with multiple types of measurements, which can be instantaneous ($J$, $K$) or of finite length in time ($H$), and corresponding time series. e) Scatter plot of the quality of the estimator $Q=\Braket{\hat{\sigma}}/\Braket{\sigma}$ for the network from a). The rates were parametrized as $k_{ij} = \kappa_{ij}e^{A_{ij}/2}$ with $\kappa_{ij}=\kappa_{ji}$ and $A_{ij}=-A_{ji}$. The rate amplitude $\kappa_{ij}$ is drawn from a uniform distribution on $[0.01, 10]$; the affinity is fixed as $A_{ii+1}=1$. A measure of the asymmetry of the network is $s=\Braket{(\kappa_{ij}-\Braket{\kappa_{ij}})^2}/\Braket{\kappa_{ij}}^2$ where the averages are taken over the three links. For the brighter dots, only the first two terms in the sum in Eq.~\eqref{eq:sigmaExample1} are considered. The dark X markers show the improvement if the first four terms in the sum in Eq.~\eqref{eq:sigmaExample1} are considered. The dashed line shows the average quality factor as a function of $s$.}
\label{fig:intro} 
\end{figure*}

\textit{Introduction and illustration of main result.---}Thermodynamic equilibrium is characterized by the absence of dissipative and irreversible processes. While dissipation is observed as heat production in the environment, the related concept of irreversibility is often summarized under the notion of an ''arrow of time''. Thus, the study of time-series of appropriate observables can reveal thermodynamic features like heat production or its conceptually more refined version, irreversibility, as quantified by entropy production. Estimation methods for entropy production are thus a central tool for inference from a thermodynamic perspective.

The most prominent class of estimators is based on simply measuring a time-asymmetric quantity like a steady state current. This concept has been developed into methods as varied as the thermodynamic uncertainty relation (TUR) \cite{bara15, ging16, horo20}, apparent entropy production rates \cite{espo12} and, in the paradigmatic case of Markov networks, fluctuation theorems for incomplete, effective descriptions \cite{shir14, pole17a, bisk17}. These methods are closely related to a second category of estimators that are based on first identifying a coarse-grained model, on which time-asymmetric currents are then identified in a second step. While older work mainly focuses on lumping Markov states together into coarse-grained ones, see, \textit{e.g.}, \cite{raha07, bo14}, much richer and more accurate descriptions are obtained by including waiting times and milestoning in semi-Markov models \cite{mart19, hartich2021, ertel2022, vdm2022, harunari2022}. Thermodynamic consistency of the resulting entropy estimators is ensured through information-theoretic \cite{cove06} reasoning, which has already found its way into stochastic thermodynamics \cite{ito20} and, in particular, into entropy estimation \cite{kawa07, gome08a, rold10}. 

Further estimation techniques are applicable by presuming a particular underlying model or equation of motion. If, for example, a master equation describes the system on some underlying level, this insight can be utilized in, \textit{e.g.}, minimization \cite{skin21, skin21a} and decimation \cite{teza2020} methods or by analyzing the communication between subsystems \cite{wolpert2020}. More generally, not only the mean but also the distribution of fluctuating entropy production is estimated and studied through application of a variety of mathematical tools including, \textit{e.g.}, large deviation theory \cite{touc09, piet15, maes17}, fluctuation-response relations \cite{hara05, dechant2020, dech21}, martingale and decision theory \cite{roldan2015, neri17, pigo17}.

From a broader perspective, ''thermodynamic inference'' \cite{seif19} is not limited to the estimation of a single quantity like entropy production. Ranging from linear systems \cite{lucente2022} over active particles \cite{ fodo16, speck2016, nard17, mand17, piet17b, dabe19, flenner2020} to living systems \cite{gnesotto2018, roldan2021}, even qualitatively distinguishing nonequilibrium from equilibrium can be challenging. In contrast, concepts like the TUR or, more recently, waiting and first passage time distributions, are not only able to infer entropy production, but also driving affinities of thermodynamic cycles \cite{bara15a, piet16, vdm2022} or topological features of transition paths \cite{satija2020, berezhkovskii2021}. 

The methods described above fall short of sufficient generality and versatility to give a nontrivial result in a model system that is as simple as the three-state network shown in Fig.~\ref{fig:intro} a) with its experimentally accessible coarse-grained version in Fig.~\ref{fig:intro} b). The coarse-grained model cannot sustain any steady state current, as it consists of two objects, $1$ and $J$, of which only the former, $1$, is a state. Since $J$ is not Markovian, we cannot employ any of the methods derived for Markov networks.

In this Letter, we introduce a framework to obtain a more general and flexible  estimator for the mean entropy production rate $\braket{\sigma}$. It remains agnostic of the underlying model and can be applied to any conceivable coarse-graining of some dynamics that can be described by a path weight, like, \textit{e.g.}, a Langevin or a master equation dynamics. In particular, the estimator is able to exploit data consisting of time-symmetric instantaneous non-Markovian events like the observation of a (non-directed) transition. For the model shown in Fig.~\ref{fig:intro}, the general entropy estimator derived below becomes
\begin{equation}
    \Braket{\hat{\sigma}} = \frac{1}{\braket{t}} \sum _{k \geq 1} \left( \prod_{j=1}^{k} \int_0^\infty dt_j \right) \psi(t_1, ..., t_{k}) \ln \frac{\psi(t_1, ..., t_{k})}{\psi(t_{k}, ..., t_1)}
    \label{eq:sigmaExample1}
,\end{equation}
where the time-resolved statistics is expressed via the  waiting time distribution $\psi(t_1, ..., t_{k})$ associated with observing $k-1$ transitions along edge $J$ with waiting times
\begin{equation}
    1 \overset{t_1}{\longrightarrow} J  \overset{t_2}{\longrightarrow} \cdots \overset{t_{k-1}}{\longrightarrow} J \overset{t_{k}}{\longrightarrow} 1
    \label{eq:bspGamma}
\end{equation}
between two visits of state $1$. The time conversion factor $1/\braket{t}$ measures the rate with which state $1$ is visited (or left) during a long trajectory. We will prove that the estimator $\braket{\hat{\sigma}}$ provides a lower bound on the total entropy production
\begin{equation}
    \Braket{\sigma} \geq \Braket{\hat{\sigma}} \geq 0
.\end{equation}

\textit{General setup.---}In the general setup, we consider coarse-grained trajectories $\Gamma$ that may contain measurements of any kind, which we will refer to as events. Fig.~\ref{fig:intro} d) shows an example. These events can be instantaneous, like $J$ and $K$, or last for a certain duration, as for $I_0$, $H$ and $I_1$. Moreover, we distinguish between two classes of measurements, which we denote by Markovian and non-Markovian events. 

If registering the event determines the state of the underlying fundamental system completely, data prior to the measurement does not contain any significant information anymore and can be disregarded. In this sense, we understand these events as Markovian. Two simple examples are the observation of a directed transition \cite{vdm2022, harunari2022} or a state in a Markov network. Markovian events allow us to cut the trajectory into smaller sections without loss of information. A section that starts and ends at a Markovian event contains the same information regardless of the remaining trajectory it is embedded in. We denote the sections that result from cutting a trajectory at every such Markovian event as ''trajectory snippets'' $\Gamma^s$. These snippets are a crucial concept of our approach. 

The time-series data obtained by observing such a coarse-grained trajectory consists of snippets of which we illustrate one in the lower part of Fig.~\ref{fig:intro} d). Along its way from the initial Markovian event $I_0$ to the final one, $I_1$, the snippet contains the events $J$, $H$ and $K$ with corresponding waiting times $t_1, ..., t_5$. We use the term ''waiting times'' in a more general way to include both genuine waiting times between consecutive events as well as residence times for events of finite duration.

\textit{Derivation of main result.---}For a stationary process, the entropy production rate in the long time limit takes the form of a Kullback-Leibler divergence after averaging \cite{kawa07, rold10},
\begin{equation}
    \Braket{\sigma} = \lim_{T \to \infty} \Braket{S}/T = \frac{1}{T} \sum_\zeta \mathcal{P}\left[\zeta\right] \ln \left( \mathcal{P}\left[\zeta\right]/\mathcal{P}[\tilde{\zeta}] \right)
,\end{equation}
where $\Braket{\cdot}$ denotes the average over many realizations $\zeta$. The time-reversal operation $\zeta \mapsto \tilde{\zeta}$ is an involution whose exact form needs to be justified by the underlying physical mechanisms. A coarse-grained trajectory $\Gamma$ is the result of a many-to-one mapping $\zeta \mapsto \Gamma(\zeta)$. This mapping determines the time-reversal operation on the coarse-grained level in terms of the underlying one, since coarse-graining has to respect $\tilde{\zeta} \mapsto \widetilde{\Gamma}$. Thus, a coarse-grained entropy production rate $\braket{\hat{\sigma}}$ can be defined, which provides an estimator for the actual entropy production in the sense that
\begin{align}
    \Braket{\sigma} \geq \Braket{\hat{\sigma}} \equiv \frac{1}{T} \sum_\Gamma \mathcal{P}\left[\Gamma\right] \ln \frac{\mathcal{P}\left[\Gamma\right]}{\mathcal{P}[\widetilde{\Gamma}]} \geq 0
    \label{eq:estimIneq}
.\end{align}
This result can be derived from well-known results in information theory like the log-sum inequality \cite{gome08a, rold10, seif19, cove06}. In this abstract form, the estimator $\braket{\hat{\sigma}}$ is both simple and universal, but often not practical, since a statistically significant amount of long trajectories with negative entropy production are needed to determine $\mathcal{P}\left[\Gamma\right]$ and $\mathcal{P}[\widetilde{\Gamma}]$. 

Instead, it is more feasible to cut $\Gamma$ into trajectory snippets of shorter length and collect the statistics of these smaller snippets. If the first (second, ...) part of the trajectory $\Gamma$ is denoted by $\Gamma_1$ ($\Gamma_2$, ...), the path weight $\mathcal{P}[\Gamma_k]$ of each individual section can be conditioned on its past in the form
\begin{align}
    \mathcal{P}[\Gamma] = \mathcal{P}[\Gamma_1]\mathcal{P}[\Gamma_2|\Gamma_1]\cdots \mathcal{P}[\Gamma_n|\Gamma_{n-1}, \Gamma_{n-2},...]
.\end{align}
If we can cut the full trajectory in such a way that the coarse-grained initial state $I_{k-1}$ of a trajectory section $\Gamma_k$ suffices to determine the state of the system on the fundamental level, the path weight factorizes into the contributions from the individual snippets in the form
\begin{align}
    \mathcal{P}[\Gamma] = \mathcal{P}[\Gamma_1^s]\mathcal{P}[\Gamma_2^s|I_1]\cdots \mathcal{P}[\Gamma_n^s|I_{n-1}]
    \label{eq:factorGamma}
.\end{align}
This condition is satisfied if we can observe at least one recurring Markovian event, which we then use as a cutting locus. Being able to consider shorter parts of the trajectories individually also emphasizes the major practical advantages that the concept of snippets offers.

In general, a snippet $\Gamma^s$ is a section of the full coarse-grained trajectory $\Gamma$. As a small trajectory on its own, a snippet is characterized by an initial state $I$, final state $J$, duration $t$ and possibly additional observations including events and waiting times, summarized under the symbol $\mathcal{O}$. The probability distribution $\psi_{I \to J}(t;\mathcal{O})$ to observe a coarse-grained trajectory $\Gamma^s$ of this form is given by the sum over all contributing microscopic trajectories $\gamma$ denoted by $\gamma_{I \to J}(t;\mathcal{O})$, \textit{i.e.},
\begin{align}
    \psi_{I \to J}(t;\mathcal{O}) \equiv \mathcal{P}[\Gamma^s|I] = \sum_{\gamma_{I \to J}(t;\mathcal{O})} \mathcal{P}[\gamma_{I \to J}(t;\mathcal{O})|I]
    \label{eq:defPsi}
\end{align}
These probability distributions can be interpreted as generalized waiting time distributions, whose normalization is written as
\begin{align}
    \sum_{J, \mathcal{O}} \int_0^\infty dt  \psi_{I \to J}(t;\mathcal{O}) = 1
.\end{align}
If the additional observations $\mathcal{O}$ contain continuous degrees of freedom, \textit{e.g.}, position in continuous space or further waiting times, the sum over $\mathcal{O}$ has to be replaced by an appropriate integral.

Up to boundary terms, using \eqref{eq:factorGamma} the coarse-grained entropy production rate \eqref{eq:estimIneq} becomes a steady state average
\begin{align}
    T \Braket{\hat{\sigma}} = \left\langle\ln \frac{\mathcal{P}[\Gamma_1^s|I_0]}{\mathcal{P}[\widetilde{\Gamma}_1^s|\tilde{I_1}]} + \cdots +  \ln \frac{\mathcal{P}[\Gamma_n^s|I_{n-1}]}{\mathcal{P}[\widetilde{\Gamma}_n^s|\tilde{I}_n]}\right\rangle
    \label{eq:sigmaHatCalc}
,\end{align}
 where $n$ is the number of snippets. Denoting the probability that a random snippet begins with $I$ by $P(I)$, we use $\mathcal{P}[\Gamma_k^s] =P(I_{k-1})\mathcal{P}[\Gamma_k^s|I_{k-1}]$ to calculate \eqref{eq:sigmaHatCalc} as
\begin{align}
    \Braket{\hat{\sigma}} = \sum_I \frac{n P(I)}{T} \left(\sum_{\Gamma^s} \mathcal{P}[\Gamma^s|I]\ln\frac{\mathcal{P}[\Gamma^s|I]}{\mathcal{P}[\widetilde{\Gamma}^s|\tilde{J}]}\right)
\end{align}
for a long, stationary trajectory. To simplify, we define $\braket{t} \equiv T/n$, which measures the average waiting time between two events that initialize a snippet. After using Eq.~\eqref{eq:defPsi} to express the path weights in terms of the waiting time distribution, we obtain our main result
\begin{equation}
    \Braket{\hat{\sigma}} = \frac{1}{\braket{t}} \sum _{IJ, \mathcal{O}} \int_0^\infty dt \pi_I \psi_{I \to J}(t;\mathcal{O}) \ln \frac{\psi_{I \to J}(t;\mathcal{O})}{\psi_{\tilde{J} \to \tilde{I}}(t;\widetilde{\mathcal{O}})} 
    \label{eq:mainResult}
\end{equation}
with $\pi_I\equiv P(I)$ and the transformations $I \mapsto \widetilde{I}, J \mapsto \widetilde{J}$ and $\mathcal{O} \mapsto \widetilde{\mathcal{O}}$ under time-reversal. As long as this behavior is known, $\mathcal{O}$ may contain any kind of events, even time-symmetric ones. In this sense, Eq.~\eqref{eq:mainResult} provides a model-free estimator without relying on particular classes of dynamics, events or model systems. We first illustrate how to apply the general Eq.~\eqref{eq:mainResult} to the paradigmatic example of Fig.~\ref{fig:intro}, which exclusively contains time-symmetric events.

\textit{Paradigmatic example.---}In the model of Fig.~\ref{fig:intro}, the coarse-grained description includes transitions along the edge $J$, whose direction is not resolved, as instantaneous non-Markovian events and a single observed Markov state $I=1$. Thus, this state is a sensible choice as initial and final event of trajectory snippets $\Gamma^s$. In other words, each snippet starts as soon as the system exits state $1$ and is terminated as soon the system revisits this state. The observations $\mathcal{O}$ along a generic trajectory snippet consist of $k-1$ transitions along the observed edge $J$ and the waiting times $t_1, ..., t_{k}$ as defined by the trajectory snippet \eqref{eq:bspGamma}. The associated waiting time distribution is given by
\begin{equation}\label{eq:intro_WTDS}
    \psi_{1 \to 1}(t;\mathcal{O}) = \psi(t_1, ..., t_{k})
,\end{equation}
with $t$ implicitly fixed via $t = \sum_{i=1}^{k} t_i$. Since both the transition $J$ and the state $1$ are even under time-reversal, the reversed trajectory is obtained by simply reading the trajectory snippet \eqref{eq:bspGamma} backwards and hence is associated with the waiting time distribution $\psi(t_k, ..., t_1)$. Therefore, the general result \eqref{eq:mainResult} reduces to Eq.~\eqref{eq:sigmaExample1}.

From a practical point of view, generating sufficient statistics for all waiting time distributions \eqref{eq:intro_WTDS} might not be feasible. Since each term in the sum in Eq.~\eqref{eq:sigmaExample1} has a non-negative contribution to $\Braket{\hat{\sigma}}$, considering only snippets that contain a maximum of $k-1$ transitions along the observed edge $J$ already results in a non-trivial estimator. The scatter plot in Fig.~\ref{fig:intro} e) illustrates this procedure. For the blue dots, only snippets with $k \leq 2$, \textit{i.e.}, $\psi(t)$ and $\psi(t_1,t_2)$, were considered. The black dots denote a random selection from these, for which we have also calculated the improvements of $\Braket{\hat{\sigma}}$ when going to $k \leq 4$, shown by the corresponding black X markers. As Fig.~\ref{fig:intro} e) also shows, the estimator inherently benefits from asymmetries in the network, which tend to produce time series with more distinct forward and backward directions.

\begin{figure}[t]
\begin{center}
    %\includestandalone[scale=1]{Tikz_snip_trans_v2.tex}
    \includegraphics[scale=1]{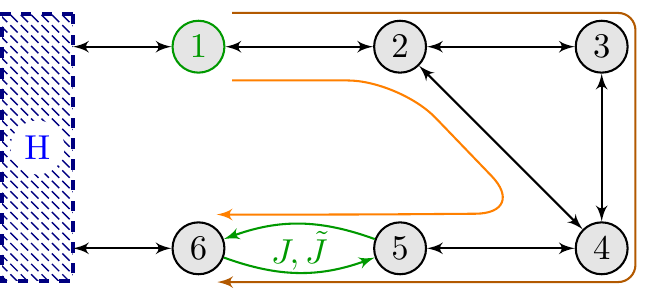}
    \caption{Illustrative example with a hidden cycle. We assume that the Markov state $1$, the directed transitions $J=(56)$ and $\tilde{J}=(65)$, and the non-Markovian event $H$ can be observed. Generically, $a_{1\to J}(t, \mathcal{O}=\emptyset)$ depends on time, which allows us to infer the existence of a hidden cycle, in this case made up by $2$, $3$ and $4$. Similarly, we are able to detect the presence of a hidden cycle within $H$ by studying the time-dependence of $a_{1 \to \tilde{J}}(t; \mathcal{O})$ for snippets that contain $H$.}
    \label{fig:sniptrans}
\end{center}
\end{figure}	

\textit{Snippets as dressed transitions.---}The time-resolved statistics is condensed in a generalized waiting time distribution $\psi_{I \to J}(t;\mathcal{O})$ that can be interpreted on the level of individual transitions $I \to J$ dressed with $t$ and $\mathcal{O}$ as additional information. The condition of apparent detailed balance on the coarse-grained level, $\pi_I \psi_{I \to J}(t;\mathcal{O}) = \pi_{\tilde{J}} \psi_{\tilde{J} \to \tilde{I}}(t;\widetilde{\mathcal{O}})$, is equivalent to \begin{equation}
    0 = \ln \frac{\pi_I \psi_{I \to J}(t;\mathcal{O})}{\pi_{\tilde{J}} \psi_{\tilde{J} \to \tilde{I}}(t;\widetilde{\mathcal{O}})} \equiv a_{I \to J}(t; \mathcal{O})
    \label{eq:defA}
,\end{equation}
which suggests to introduce $a_{I \to J}(t; \mathcal{O})$ as a thermodynamic measure of irreversibility inherent to the dressed transition $I \to J$. Thus, if at least one $a_{I \to J}(t; \mathcal{O})$ does not vanish, the system cannot be in equilibrium. Moreover, the events $I, J$ and those included in $\mathcal{O}$ have to be part of a thermodynamic cycle with nonvanishing affinity if $a_{I \to J}(t; \mathcal{O}) \neq 0$. 

Surprisingly, we are even able to infer hidden cycles that do not necessarily include the observed events. We illustrate this method in Fig.~\ref{fig:sniptrans}. We observe an explicit time-dependence of $a_{1 \to J}(t,\mathcal{O}=\emptyset)$, \textit{i.e.}, broken local detailed balance \cite{seif19} in the trajectory snippets from $1$ to $J$, if and only if the affinity of the small cycle containing the states $2,3,4$ does not vanish. This broken symmetry between forward and backward transitions generalizes extant discussions for Markov networks \cite{berezhkovskii2019} and Langevin dynamics \cite{berezhkovskii2006}.

\begin{figure}[t]
\begin{center}
    %\includestandalone[scale=0.34375]{nj_plot.tex}
    \includegraphics[scale=0.34375]{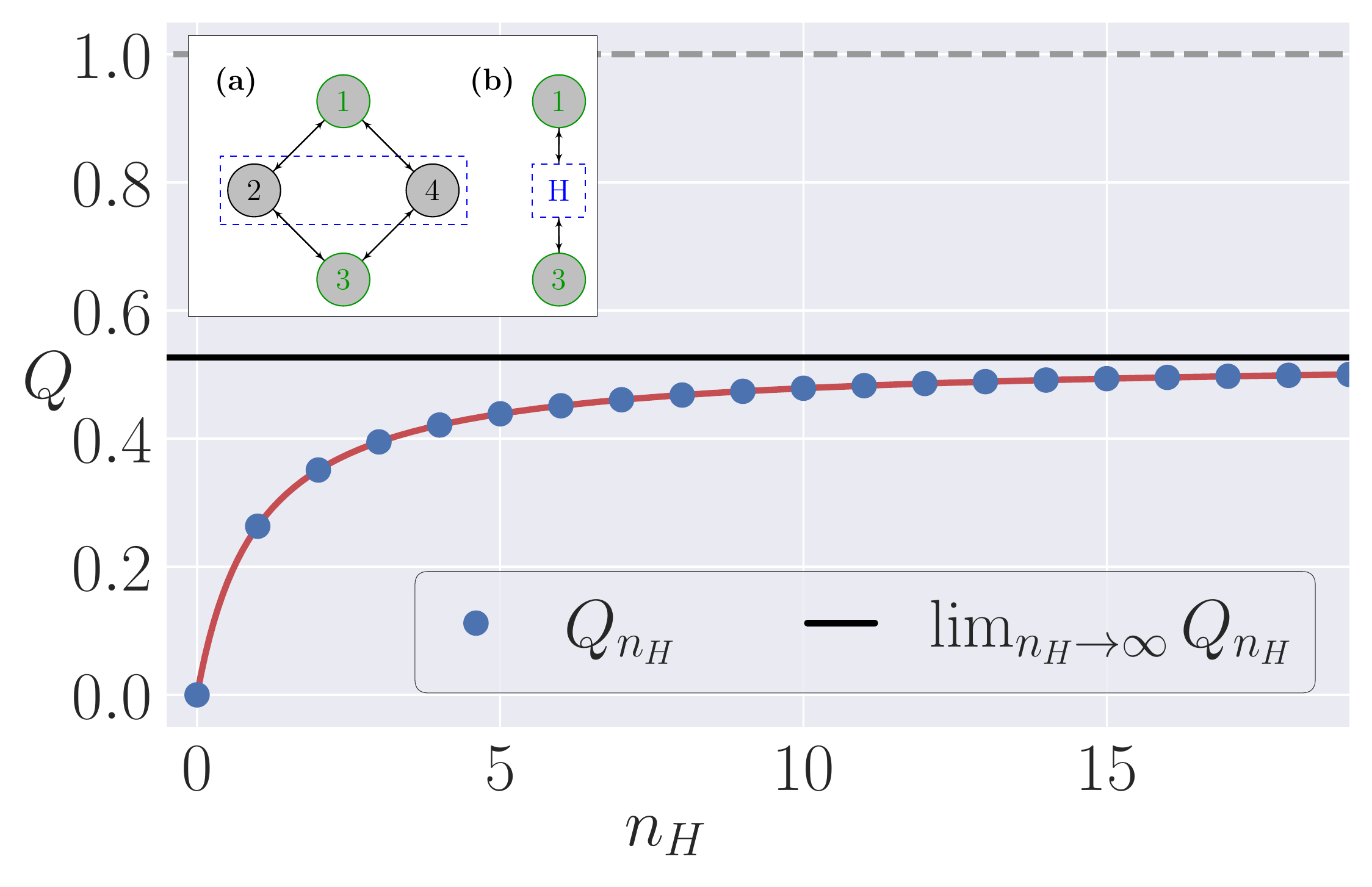}
    \caption{Demonstration of the systematic bias in the entropy estimator if the trajectory is cut at a non-Markovian state $H$.  The inset shows the full (a) and coarse-grained (b) network.  $Q_{n_H} \equiv \braket{\hat{\sigma}_{n_H}}/\braket{\sigma}$ results from cutting the trajectory at every $n_H$-th occurrence of $H$. For $n_H\to\infty$ it converges to the estimator obtained by cutting the trajectory at the Markov states $1$ and $3$. The rates are $k_{12}=3$, $k_{21}=1$, $k_{23}=9$, $k_{32}=3$, $k_{34}=27$, $k_{43}=9$, $k_{41}=81$ and $k_{14}=27$.}
    \label{fig:nj}
\end{center}
\end{figure}	

\textit{Non-Markovian snippets.---}In the previous sections, a Markov property in the form of Eq.~\eqref{eq:factorGamma} has turned out to be crucial. What happens if we cut a trajectory into snippets at, say, a hidden compound state where this factorization is not valid? Writing the entropy production rate as \cite{rold10}
\begin{equation}
    \Braket{\sigma} = \lim_{T \to \infty} \frac{1}{T} \sum_{\gamma_T} \mathcal{P}[\gamma_T] \ln \frac{\mathcal{P}[\gamma_T]}{\mathcal{P}[\tilde{\gamma}_T]}
\end{equation}
for trajectories $\gamma_T$ of length $T$ suggests asymptotic consistency of an entropy estimator of the form \eqref{eq:mainResult} if the length $t$ of snippets become sufficiently large on average. Conditioning the waiting-time distribution $\psi_{H \to J}(t;\mathcal{O})$ on a non-Markovian event $H$ while still disregarding the past of the trajectories introduces a systematic bias. This bias can be used as an operationally verifiable, model-independent criterion for Markovianity of some measured event $H$. If the observed trajectory $\Gamma$ is cut at, say, every $n_H$-th occurrence of $H$ rather than every occurrence of $H$, the implied entropy estimator $\braket{\hat{\sigma}_{n_H}}$ is independent of $n_H$ if and only if the factorization \eqref{eq:factorGamma} can be applied, \textit{i.e.}, if $H$ qualifies as a cut locus for snippets. As an example, Fig.~\ref{fig:nj} shows a four state Markov network where states $2$ and $4$ form a compound state $H$. Non-Markovian snippets obtained by cutting at every $n_H$-th occurrence of $H$ lead to an estimator, which depends on $n_H$ and improves for $n_H\to\infty$.

\begin{figure}[t]
\centering
%\includestandalone[scale=1]{Tikz_sigma_overest.tex}
\includegraphics[scale=1]{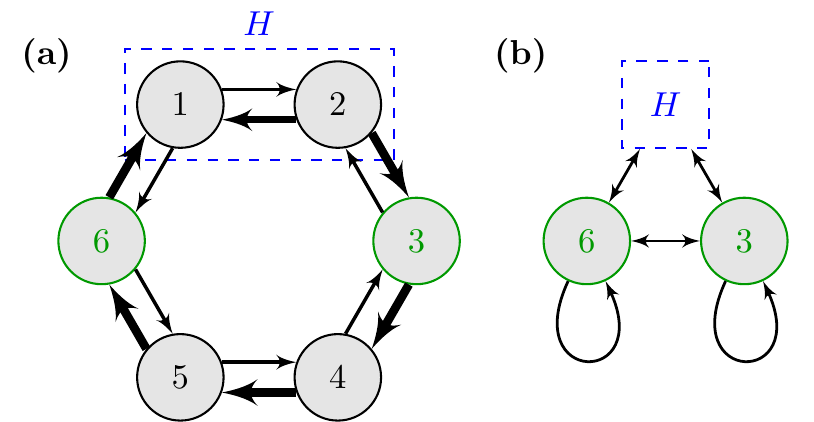}
\caption{Illustration of a system (a) and its coarse-grained description (b). The width of the arrows in (a) indicates the magnitude of the rates. We assume that only the Markov states $3$ and $6$ and the compound state $H$ can be observed. Treating $H$ as a Markov state, \textit{i.e.}, as locus for cutting trajectories, can lead to an overestimation of the cycle affinity and entropy production.}
\label{fig:sigmaoverest} 
\end{figure}

Thus, shorter snippets reduce the statistical error of finite sample sizes at the cost of introducing systematic bias. This systematic bias may even overestimate $\Braket{\sigma}$, \textit{i.e.}, the error is qualitatively different from merely disregarding information, which always decreases entropy production. Fig.~\ref{fig:sigmaoverest} presents an example where this systematic bias can lead to an overestimation of the entropy production. 

\textit{Concluding perspective.---}This work has established a framework for constructing an entropy estimator based on coarse-grained data, which may include any kind of measurement whose behavior under time reversal is known and the associated time-resolved statistics. While we have assumed a stationary process, \textit{i.e.}, a non-equilibrium steady state, the underlying concepts are sufficiently general so that future work can adapt this approach to time-dependent situations like periodic driving. 

Moreover, we want to emphasize the shift in focus regarding the Markov property. While Markovianity is usually understood as a system property, \textit{e.g.}, in the case of overdamped Langevin dynamics or Markov networks, the present formalism is based on Markovianity as a property of particular observable events. In accordance with Occam's razor, we do not make any assumptions about unobservable parts of the system, which renders this approach more applicable to model complex real-world scenarios.

Lastly, our work also provides a starting point for thermodynamic inference beyond the estimation of a single quantity like entropy production. We have pointed out how broken local detailed balance can be detected qualitatively. If more details about the system are known, these concepts can be quantified into estimators for driving affinity and topology of the thermodynamic cycles as Ref. \cite{vdm2022} has shown for trajectories cut at observed directed transitions. The much broader framework developed here invites further advances in this direction of a more refined thermodynamic inference scheme.

\textit{Acknowledgments.---}We thank Benjamin Ertel for valuable discussions.

\bibliography{speedrun, references}

\begin{thebibliography}{49}%
\makeatletter
\providecommand \@ifxundefined [1]{%
 \@ifx{#1\undefined}
}%
\providecommand \@ifnum [1]{%
 \ifnum #1\expandafter \@firstoftwo
 \else \expandafter \@secondoftwo
 \fi
}%
\providecommand \@ifx [1]{%
 \ifx #1\expandafter \@firstoftwo
 \else \expandafter \@secondoftwo
 \fi
}%
\providecommand \natexlab [1]{#1}%
\providecommand \enquote  [1]{``#1''}%
\providecommand \bibnamefont  [1]{#1}%
\providecommand \bibfnamefont [1]{#1}%
\providecommand \citenamefont [1]{#1}%
\providecommand \href@noop [0]{\@secondoftwo}%
\providecommand \href [0]{\begingroup \@sanitize@url \@href}%
\providecommand \@href[1]{\@@startlink{#1}\@@href}%
\providecommand \@@href[1]{\endgroup#1\@@endlink}%
\providecommand \@sanitize@url [0]{\catcode `\\12\catcode `\$12\catcode
  `\&12\catcode `\#12\catcode `\^12\catcode `\_12\catcode `\%12\relax}%
\providecommand \@@startlink[1]{}%
\providecommand \@@endlink[0]{}%
\providecommand \url  [0]{\begingroup\@sanitize@url \@url }%
\providecommand \@url [1]{\endgroup\@href {#1}{\urlprefix }}%
\providecommand \urlprefix  [0]{URL }%
\providecommand \Eprint [0]{\href }%
\providecommand \doibase [0]{https://doi.org/}%
\providecommand \selectlanguage [0]{\@gobble}%
\providecommand \bibinfo  [0]{\@secondoftwo}%
\providecommand \bibfield  [0]{\@secondoftwo}%
\providecommand \translation [1]{[#1]}%
\providecommand \BibitemOpen [0]{}%
\providecommand \bibitemStop [0]{}%
\providecommand \bibitemNoStop [0]{.\EOS\space}%
\providecommand \EOS [0]{\spacefactor3000\relax}%
\providecommand \BibitemShut  [1]{\csname bibitem#1\endcsname}%
\let\auto@bib@innerbib\@empty
%</preamble>
\bibitem [{\citenamefont {Barato}\ and\ \citenamefont
  {Seifert}(2015{\natexlab{a}})}]{bara15}%
  \BibitemOpen
  \bibfield  {author} {\bibinfo {author} {\bibfnamefont {A.~C.}\ \bibnamefont
  {Barato}}\ and\ \bibinfo {author} {\bibfnamefont {U.}~\bibnamefont
  {Seifert}},\ }\bibfield  {title} {\bibinfo {title} {Thermodynamic uncertainty
  relation for biomolecular processes},\ }\href
  {https://doi.org/10.1103/PhysRevLett.114.158101} {\bibfield  {journal}
  {\bibinfo  {journal} {Phys. Rev. Lett.}\ }\textbf {\bibinfo {volume} {114}},\
  \bibinfo {pages} {158101} (\bibinfo {year} {2015}{\natexlab{a}})}\BibitemShut
  {NoStop}%
\bibitem [{\citenamefont {Gingrich}\ \emph {et~al.}(2016)\citenamefont
  {Gingrich}, \citenamefont {Horowitz}, \citenamefont {Perunov},\ and\
  \citenamefont {England}}]{ging16}%
  \BibitemOpen
  \bibfield  {author} {\bibinfo {author} {\bibfnamefont {T.~R.}\ \bibnamefont
  {Gingrich}}, \bibinfo {author} {\bibfnamefont {J.~M.}\ \bibnamefont
  {Horowitz}}, \bibinfo {author} {\bibfnamefont {N.}~\bibnamefont {Perunov}},\
  and\ \bibinfo {author} {\bibfnamefont {J.~L.}\ \bibnamefont {England}},\
  }\bibfield  {title} {\bibinfo {title} {Dissipation bounds all steady-state
  current fluctuations},\ }\href
  {https://doi.org/10.1103/PhysRevLett.116.120601} {\bibfield  {journal}
  {\bibinfo  {journal} {Phys. Rev. Lett.}\ }\textbf {\bibinfo {volume} {116}},\
  \bibinfo {pages} {120601} (\bibinfo {year} {2016})}\BibitemShut {NoStop}%
\bibitem [{\citenamefont {Horowitz}\ and\ \citenamefont
  {Gingrich}(2020)}]{horo20}%
  \BibitemOpen
  \bibfield  {author} {\bibinfo {author} {\bibfnamefont {J.~M.}\ \bibnamefont
  {Horowitz}}\ and\ \bibinfo {author} {\bibfnamefont {T.~R.}\ \bibnamefont
  {Gingrich}},\ }\bibfield  {title} {\bibinfo {title} {Thermodynamic
  uncertainty relations constrain non-equilibrium fluctuations},\ }\href
  {https://doi.org/10.1038/s41567-019-0702-6} {\bibfield  {journal} {\bibinfo
  {journal} {Nat. Phys.}\ }\textbf {\bibinfo {volume} {16}},\ \bibinfo {pages}
  {15} (\bibinfo {year} {2020})}\BibitemShut {NoStop}%
\bibitem [{\citenamefont {Esposito}(2012)}]{espo12}%
  \BibitemOpen
  \bibfield  {author} {\bibinfo {author} {\bibfnamefont {M.}~\bibnamefont
  {Esposito}},\ }\bibfield  {title} {\bibinfo {title} {Stochastic
  thermodynamics under coarse-graining},\ }\href
  {https://doi.org/10.1103/PhysRevE.85.041125} {\bibfield  {journal} {\bibinfo
  {journal} {Phys. Rev. E}\ }\textbf {\bibinfo {volume} {85}},\ \bibinfo
  {pages} {041125} (\bibinfo {year} {2012})}\BibitemShut {NoStop}%
\bibitem [{\citenamefont {Shiraishi}\ and\ \citenamefont
  {Sagawa}(2015)}]{shir14}%
  \BibitemOpen
  \bibfield  {author} {\bibinfo {author} {\bibfnamefont {N.}~\bibnamefont
  {Shiraishi}}\ and\ \bibinfo {author} {\bibfnamefont {T.}~\bibnamefont
  {Sagawa}},\ }\bibfield  {title} {\bibinfo {title} {Fluctuation theorem for
  partially masked nonequilibrium dynamics},\ }\href
  {https://doi.org/10.1103/PhysRevE.91.012130} {\bibfield  {journal} {\bibinfo
  {journal} {Phys. Rev. E}\ }\textbf {\bibinfo {volume} {91}},\ \bibinfo
  {pages} {012130} (\bibinfo {year} {2015})}\BibitemShut {NoStop}%
\bibitem [{\citenamefont {Polettini}\ and\ \citenamefont
  {Esposito}(2017)}]{pole17a}%
  \BibitemOpen
  \bibfield  {author} {\bibinfo {author} {\bibfnamefont {M.}~\bibnamefont
  {Polettini}}\ and\ \bibinfo {author} {\bibfnamefont {M.}~\bibnamefont
  {Esposito}},\ }\bibfield  {title} {\bibinfo {title} {Effective thermodynamics
  for a marginal observer},\ }\href
  {https://doi.org/10.1103/PhysRevLett.119.240601} {\bibfield  {journal}
  {\bibinfo  {journal} {Phys. Rev. Lett.}\ }\textbf {\bibinfo {volume} {119}},\
  \bibinfo {pages} {240601} (\bibinfo {year} {2017})}\BibitemShut {NoStop}%
\bibitem [{\citenamefont {Bisker}\ \emph {et~al.}(2017)\citenamefont {Bisker},
  \citenamefont {Polettini}, \citenamefont {Gingrich},\ and\ \citenamefont
  {Horowitz}}]{bisk17}%
  \BibitemOpen
  \bibfield  {author} {\bibinfo {author} {\bibfnamefont {G.}~\bibnamefont
  {Bisker}}, \bibinfo {author} {\bibfnamefont {M.}~\bibnamefont {Polettini}},
  \bibinfo {author} {\bibfnamefont {T.~R.}\ \bibnamefont {Gingrich}},\ and\
  \bibinfo {author} {\bibfnamefont {J.~M.}\ \bibnamefont {Horowitz}},\
  }\bibfield  {title} {\bibinfo {title} {Hierarchical bounds on entropy
  production inferred from partial information},\ }\href
  {https://doi.org/10.1088/1742-5468/aa8c0d} {\bibfield  {journal} {\bibinfo
  {journal} {J. Stat. Mech. Theor. Exp.}\ }\textbf {\bibinfo {volume} {2017}},\
  \bibinfo {pages} {093210} (\bibinfo {year} {2017})}\BibitemShut {NoStop}%
\bibitem [{\citenamefont {Rahav}\ and\ \citenamefont
  {Jarzynski}(2007)}]{raha07}%
  \BibitemOpen
  \bibfield  {author} {\bibinfo {author} {\bibfnamefont {S.}~\bibnamefont
  {Rahav}}\ and\ \bibinfo {author} {\bibfnamefont {C.}~\bibnamefont
  {Jarzynski}},\ }\bibfield  {title} {\bibinfo {title} {Fluctuation relations
  and coarse-graining},\ }\href
  {https://doi.org/10.1088/1742-5468/2007/09/P09012} {\bibfield  {journal}
  {\bibinfo  {journal} {J. Stat. Mech.: Theor. Exp.}\ }\textbf {\bibinfo
  {volume} {2007}},\ \bibinfo {pages} {P09012} (\bibinfo {year}
  {2007})}\BibitemShut {NoStop}%
\bibitem [{\citenamefont {Bo}\ and\ \citenamefont {Celani}(2014)}]{bo14}%
  \BibitemOpen
  \bibfield  {author} {\bibinfo {author} {\bibfnamefont {S.}~\bibnamefont
  {Bo}}\ and\ \bibinfo {author} {\bibfnamefont {A.}~\bibnamefont {Celani}},\
  }\bibfield  {title} {\bibinfo {title} {Entropy production in stochastic
  systems with fast and slow time-scales},\ }\href
  {https://doi.org/10.1007/s10955-014-0922-1} {\bibfield  {journal} {\bibinfo
  {journal} {J. Stat. Phys.}\ }\textbf {\bibinfo {volume} {154}},\ \bibinfo
  {pages} {1325} (\bibinfo {year} {2014})}\BibitemShut {NoStop}%
\bibitem [{\citenamefont {Martinez}\ \emph {et~al.}(2019)\citenamefont
  {Martinez}, \citenamefont {Bisker}, \citenamefont {Horowitz},\ and\
  \citenamefont {Parrondo}}]{mart19}%
  \BibitemOpen
  \bibfield  {author} {\bibinfo {author} {\bibfnamefont {I.~A.}\ \bibnamefont
  {Martinez}}, \bibinfo {author} {\bibfnamefont {G.}~\bibnamefont {Bisker}},
  \bibinfo {author} {\bibfnamefont {J.~M.}\ \bibnamefont {Horowitz}},\ and\
  \bibinfo {author} {\bibfnamefont {J.~M.~R.}\ \bibnamefont {Parrondo}},\
  }\bibfield  {title} {\bibinfo {title} {Inferring broken detailed balance in
  the absence of observable currents},\ }\href
  {https://doi.org/10.1038/s41467-019-11051-w} {\bibfield  {journal} {\bibinfo
  {journal} {Nature Communications}\ }\textbf {\bibinfo {volume} {10}},\
  \bibinfo {pages} {3542} (\bibinfo {year} {2019})}\BibitemShut {NoStop}%
\bibitem [{\citenamefont {Hartich}\ and\ \citenamefont
  {Godec}(2021)}]{hartich2021}%
  \BibitemOpen
  \bibfield  {author} {\bibinfo {author} {\bibfnamefont {D.}~\bibnamefont
  {Hartich}}\ and\ \bibinfo {author} {\bibfnamefont {A.}~\bibnamefont
  {Godec}},\ }\bibfield  {title} {{\selectlanguage {en}\bibinfo {title}
  {Emergent {Memory} and {Kinetic} {Hysteresis} in {Strongly} {Driven}
  {Networks}}},\ }\href {https://doi.org/10.1103/PhysRevX.11.041047} {\bibfield
   {journal} {\bibinfo  {journal} {Physical Review X}\ }\textbf {\bibinfo
  {volume} {11}},\ \bibinfo {pages} {041047} (\bibinfo {year}
  {2021})}\BibitemShut {NoStop}%
\bibitem [{\citenamefont {Ertel}\ \emph {et~al.}(2022)\citenamefont {Ertel},
  \citenamefont {van~der Meer},\ and\ \citenamefont {Seifert}}]{ertel2022}%
  \BibitemOpen
  \bibfield  {author} {\bibinfo {author} {\bibfnamefont {B.}~\bibnamefont
  {Ertel}}, \bibinfo {author} {\bibfnamefont {J.}~\bibnamefont {van~der
  Meer}},\ and\ \bibinfo {author} {\bibfnamefont {U.}~\bibnamefont {Seifert}},\
  }\bibfield  {title} {\bibinfo {title} {Operationally accessible uncertainty
  relations for thermodynamically consistent semi-markov processes},\ }\href
  {https://doi.org/10.1103/PhysRevE.105.044113} {\bibfield  {journal} {\bibinfo
   {journal} {Phys. Rev. E}\ }\textbf {\bibinfo {volume} {105}},\ \bibinfo
  {pages} {044113} (\bibinfo {year} {2022})}\BibitemShut {NoStop}%
\bibitem [{\citenamefont {van~der Meer}\ \emph {et~al.}(2022)\citenamefont
  {van~der Meer}, \citenamefont {Ertel},\ and\ \citenamefont
  {Seifert}}]{vdm2022}%
  \BibitemOpen
  \bibfield  {author} {\bibinfo {author} {\bibfnamefont {J.}~\bibnamefont
  {van~der Meer}}, \bibinfo {author} {\bibfnamefont {B.}~\bibnamefont
  {Ertel}},\ and\ \bibinfo {author} {\bibfnamefont {U.}~\bibnamefont
  {Seifert}},\ }\bibfield  {title} {{\selectlanguage {en}\bibinfo {title}
  {Thermodynamic {Inference} in {Partially} {Accessible} {Markov} {Networks}:
  {A} {Unifying} {Perspective} from {Transition}-{Based} {Waiting} {Time}
  {Distributions}}},\ }\href {https://doi.org/10.1103/PhysRevX.12.031025}
  {\bibfield  {journal} {\bibinfo  {journal} {Physical Review X}\ }\textbf
  {\bibinfo {volume} {12}},\ \bibinfo {pages} {031025} (\bibinfo {year}
  {2022})}\BibitemShut {NoStop}%
\bibitem [{\citenamefont {Harunari}\ \emph {et~al.}(2022)\citenamefont
  {Harunari}, \citenamefont {Dutta}, \citenamefont {Polettini},\ and\
  \citenamefont {Roldan}}]{harunari2022}%
  \BibitemOpen
  \bibfield  {author} {\bibinfo {author} {\bibfnamefont {P.~E.}\ \bibnamefont
  {Harunari}}, \bibinfo {author} {\bibfnamefont {A.}~\bibnamefont {Dutta}},
  \bibinfo {author} {\bibfnamefont {M.}~\bibnamefont {Polettini}},\ and\
  \bibinfo {author} {\bibfnamefont {E.}~\bibnamefont {Roldan}},\ }\bibfield
  {title} {{\selectlanguage {en}\bibinfo {title} {What to learn from few
  visible transitions' statistics?}},\ }\href {http://arxiv.org/abs/2203.07427}
  {\bibfield  {journal} {\bibinfo  {journal} {arXiv:2203.07427 [cond-mat,
  physics:physics]}\ } (\bibinfo {year} {2022})},\ \bibinfo {note} {arXiv:
  2203.07427}\BibitemShut {NoStop}%
\bibitem [{\citenamefont {Cover}\ and\ \citenamefont {Thomas}(2006)}]{cove06}%
  \BibitemOpen
  \bibfield  {author} {\bibinfo {author} {\bibfnamefont {T.~M.}\ \bibnamefont
  {Cover}}\ and\ \bibinfo {author} {\bibfnamefont {J.~A.}\ \bibnamefont
  {Thomas}},\ }\href@noop {} {\emph {\bibinfo {title} {Elements of Information
  Theory}}},\ Telecommunications and Signal Processing\ (\bibinfo  {publisher}
  {{Wiley}},\ \bibinfo {address} {{Hoboken, NJ, and Canada}},\ \bibinfo {year}
  {2006})\BibitemShut {NoStop}%
\bibitem [{\citenamefont {Ito}\ and\ \citenamefont {Dechant}(2020)}]{ito20}%
  \BibitemOpen
  \bibfield  {author} {\bibinfo {author} {\bibfnamefont {S.}~\bibnamefont
  {Ito}}\ and\ \bibinfo {author} {\bibfnamefont {A.}~\bibnamefont {Dechant}},\
  }\bibfield  {title} {\bibinfo {title} {Stochastic time evolution, information
  geometry, and the {Cram\'er-Rao} bound},\ }\href
  {https://doi.org/10.1103/PhysRevX.10.021056} {\bibfield  {journal} {\bibinfo
  {journal} {Phys. Rev. X}\ }\textbf {\bibinfo {volume} {10}},\ \bibinfo
  {pages} {021056} (\bibinfo {year} {2020})}\BibitemShut {NoStop}%
\bibitem [{\citenamefont {Kawai}\ \emph {et~al.}(2007)\citenamefont {Kawai},
  \citenamefont {Parrondo},\ and\ \citenamefont {{van den Broeck}}}]{kawa07}%
  \BibitemOpen
  \bibfield  {author} {\bibinfo {author} {\bibfnamefont {R.}~\bibnamefont
  {Kawai}}, \bibinfo {author} {\bibfnamefont {J.~M.~R.}\ \bibnamefont
  {Parrondo}},\ and\ \bibinfo {author} {\bibfnamefont {C.}~\bibnamefont {{van
  den Broeck}}},\ }\bibfield  {title} {\bibinfo {title} {Dissipation: {{The}}
  phase-space perspective},\ }\href
  {https://doi.org/10.1103/PhysRevLett.98.080602} {\bibfield  {journal}
  {\bibinfo  {journal} {Phys. Rev. Lett.}\ }\textbf {\bibinfo {volume} {98}},\
  \bibinfo {pages} {080602} (\bibinfo {year} {2007})}\BibitemShut {NoStop}%
\bibitem [{\citenamefont {{Gomez-Marin}}\ \emph {et~al.}(2008)\citenamefont
  {{Gomez-Marin}}, \citenamefont {Parrondo},\ and\ \citenamefont {van~den
  Broeck}}]{gome08a}%
  \BibitemOpen
  \bibfield  {author} {\bibinfo {author} {\bibfnamefont {A.}~\bibnamefont
  {{Gomez-Marin}}}, \bibinfo {author} {\bibfnamefont {J.~M.~R.}\ \bibnamefont
  {Parrondo}},\ and\ \bibinfo {author} {\bibfnamefont {C.}~\bibnamefont {van~den
  Broeck}},\ }\bibfield  {title} {\bibinfo {title} {Lower bounds on dissipation
  upon coarse-graining},\ }\href {https://doi.org/10.1103/PhysRevE.78.011107}
  {\bibfield  {journal} {\bibinfo  {journal} {Phys. Rev. E}\ }\textbf {\bibinfo
  {volume} {78}},\ \bibinfo {pages} {011107} (\bibinfo {year}
  {2008})}\BibitemShut {NoStop}%
\bibitem [{\citenamefont {Roldan}\ and\ \citenamefont
  {Parrondo}(2010)}]{rold10}%
  \BibitemOpen
  \bibfield  {author} {\bibinfo {author} {\bibfnamefont {E.}~\bibnamefont
  {Roldan}}\ and\ \bibinfo {author} {\bibfnamefont {J.~M.~R.}\ \bibnamefont
  {Parrondo}},\ }\bibfield  {title} {\bibinfo {title} {Estimating dissipation
  from single stationary trajectories},\ }\href
  {https://doi.org/10.1103/PhysRevLett.105.150607} {\bibfield  {journal}
  {\bibinfo  {journal} {Phys. Rev. Lett.}\ }\textbf {\bibinfo {volume} {105}},\
  \bibinfo {pages} {150607} (\bibinfo {year} {2010})}\BibitemShut {NoStop}%
\bibitem [{\citenamefont {Skinner}\ and\ \citenamefont
  {Dunkel}(2021{\natexlab{a}})}]{skin21}%
  \BibitemOpen
  \bibfield  {author} {\bibinfo {author} {\bibfnamefont {D.~J.}\ \bibnamefont
  {Skinner}}\ and\ \bibinfo {author} {\bibfnamefont {J.}~\bibnamefont
  {Dunkel}},\ }\bibfield  {title} {\bibinfo {title} {Improved bounds on entropy
  production in living systems},\ }\href@noop {} {\bibfield  {journal}
  {\bibinfo  {journal} {Proc. Natl. Acad. Sci. U.S.A.}\ }\textbf {\bibinfo
  {volume} {118}} (\bibinfo {year} {2021}{\natexlab{a}})}\BibitemShut {NoStop}%
\bibitem [{\citenamefont {Skinner}\ and\ \citenamefont
  {Dunkel}(2021{\natexlab{b}})}]{skin21a}%
  \BibitemOpen
  \bibfield  {author} {\bibinfo {author} {\bibfnamefont {D.~J.}\ \bibnamefont
  {Skinner}}\ and\ \bibinfo {author} {\bibfnamefont {J.}~\bibnamefont
  {Dunkel}},\ }\bibfield  {title} {\bibinfo {title} {Estimating entropy
  production from waiting time distributions},\ }\href@noop {} {\bibfield
  {journal} {\bibinfo  {journal} {Phys. Rev. Lett.}\ }\textbf {\bibinfo
  {volume} {127}},\ \bibinfo {pages} {198101} (\bibinfo {year}
  {2021}{\natexlab{b}})}\BibitemShut {NoStop}%
\bibitem [{\citenamefont {Teza}\ and\ \citenamefont {Stella}(2020)}]{teza2020}%
  \BibitemOpen
  \bibfield  {author} {\bibinfo {author} {\bibfnamefont {G.}~\bibnamefont
  {Teza}}\ and\ \bibinfo {author} {\bibfnamefont {A.~L.}\ \bibnamefont
  {Stella}},\ }\bibfield  {title} {{\selectlanguage {en}\bibinfo {title} {Exact
  {Coarse} {Graining} {Preserves} {Entropy} {Production} out of
  {Equilibrium}}},\ }\href {https://doi.org/10.1103/PhysRevLett.125.110601}
  {\bibfield  {journal} {\bibinfo  {journal} {Physical Review Letters}\
  }\textbf {\bibinfo {volume} {125}},\ \bibinfo {pages} {110601} (\bibinfo
  {year} {2020})}\BibitemShut {NoStop}%
\bibitem [{\citenamefont {Wolpert}(2020)}]{wolpert2020}%
  \BibitemOpen
  \bibfield  {author} {\bibinfo {author} {\bibfnamefont {D.~H.}\ \bibnamefont
  {Wolpert}},\ }\bibfield  {title} {{\selectlanguage {en}\bibinfo {title}
  {Minimal entropy production rate of interacting systems}},\ }\href
  {https://doi.org/10.1088/1367-2630/abc5c6} {\bibfield  {journal} {\bibinfo
  {journal} {New Journal of Physics}\ }\textbf {\bibinfo {volume} {22}},\
  \bibinfo {pages} {113013} (\bibinfo {year} {2020})}\BibitemShut {NoStop}%
\bibitem [{\citenamefont {Touchette}(2009)}]{touc09}%
  \BibitemOpen
  \bibfield  {author} {\bibinfo {author} {\bibfnamefont {H.}~\bibnamefont
  {Touchette}},\ }\bibfield  {title} {\bibinfo {title} {The large deviation
  approach to statistical mechanics},\ }\href
  {https://doi.org/10.1016/j.physrep.2009.05.002} {\bibfield  {journal}
  {\bibinfo  {journal} {Phys. Rep.}\ }\textbf {\bibinfo {volume} {478}},\
  \bibinfo {pages} {1} (\bibinfo {year} {2009})}\BibitemShut {NoStop}%
\bibitem [{\citenamefont {Pietzonka}\ \emph
  {et~al.}(2016{\natexlab{a}})\citenamefont {Pietzonka}, \citenamefont
  {Barato},\ and\ \citenamefont {Seifert}}]{piet15}%
  \BibitemOpen
  \bibfield  {author} {\bibinfo {author} {\bibfnamefont {P.}~\bibnamefont
  {Pietzonka}}, \bibinfo {author} {\bibfnamefont {A.~C.}\ \bibnamefont
  {Barato}},\ and\ \bibinfo {author} {\bibfnamefont {U.}~\bibnamefont
  {Seifert}},\ }\bibfield  {title} {\bibinfo {title} {Universal bounds on
  current fluctuations},\ }\href {https://doi.org/10.1103/PhysRevE.93.052145}
  {\bibfield  {journal} {\bibinfo  {journal} {Phys. Rev. E}\ }\textbf {\bibinfo
  {volume} {93}},\ \bibinfo {pages} {052145} (\bibinfo {year}
  {2016}{\natexlab{a}})}\BibitemShut {NoStop}%
\bibitem [{\citenamefont {Maes}(2017)}]{maes17}%
  \BibitemOpen
  \bibfield  {author} {\bibinfo {author} {\bibfnamefont {C.}~\bibnamefont
  {Maes}},\ }\bibfield  {title} {\bibinfo {title} {Frenetic bounds on the
  entropy production},\ }\href {https://doi.org/10.1103/PhysRevLett.119.160601}
  {\bibfield  {journal} {\bibinfo  {journal} {Phys. Rev. Lett.}\ }\textbf
  {\bibinfo {volume} {119}},\ \bibinfo {pages} {160601} (\bibinfo {year}
  {2017})}\BibitemShut {NoStop}%
\bibitem [{\citenamefont {Harada}\ and\ \citenamefont {Sasa}(2005)}]{hara05}%
  \BibitemOpen
  \bibfield  {author} {\bibinfo {author} {\bibfnamefont {T.}~\bibnamefont
  {Harada}}\ and\ \bibinfo {author} {\bibfnamefont {S.~I.}\ \bibnamefont
  {Sasa}},\ }\bibfield  {title} {\bibinfo {title} {Equality connecting energy
  dissipation with a violation of the fluctuation-response relation},\ }\href
  {https://doi.org/10.1103/PhysRevLett.95.130602} {\bibfield  {journal}
  {\bibinfo  {journal} {Phys. Rev. Lett.}\ }\textbf {\bibinfo {volume} {95}},\
  \bibinfo {pages} {130602} (\bibinfo {year} {2005})}\BibitemShut {NoStop}%
\bibitem [{\citenamefont {Dechant}\ and\ \citenamefont
  {Sasa}(2020)}]{dechant2020}%
  \BibitemOpen
  \bibfield  {author} {\bibinfo {author} {\bibfnamefont {A.}~\bibnamefont
  {Dechant}}\ and\ \bibinfo {author} {\bibfnamefont {S.-i.}\ \bibnamefont
  {Sasa}},\ }\bibfield  {title} {{\selectlanguage {en}\bibinfo {title}
  {Fluctuation–response inequality out of equilibrium}},\ }\href
  {https://doi.org/10.1073/pnas.1918386117} {\bibfield  {journal} {\bibinfo
  {journal} {Proceedings of the National Academy of Sciences}\ }\textbf
  {\bibinfo {volume} {117}},\ \bibinfo {pages} {6430} (\bibinfo {year}
  {2020})}\BibitemShut {NoStop}%
\bibitem [{\citenamefont {Dechant}\ and\ \citenamefont {Sasa}(2021)}]{dech21}%
  \BibitemOpen
  \bibfield  {author} {\bibinfo {author} {\bibfnamefont {A.}~\bibnamefont
  {Dechant}}\ and\ \bibinfo {author} {\bibfnamefont {S.~I.}\ \bibnamefont
  {Sasa}},\ }\bibfield  {title} {\bibinfo {title} {Improving thermodynamic
  bounds using correlations},\ }\href
  {https://doi.org/10.1103/PhysRevX.11.041061} {\bibfield  {journal} {\bibinfo
  {journal} {Phys. Rev. X}\ }\textbf {\bibinfo {volume} {11}},\ \bibinfo
  {pages} {041061} (\bibinfo {year} {2021})}\BibitemShut {NoStop}%
\bibitem [{\citenamefont {Roldan}\ \emph {et~al.}(2015)\citenamefont {Roldan},
  \citenamefont {Neri}, \citenamefont {Dörpinghaus}, \citenamefont {Meyr},\
  and\ \citenamefont {Jülicher}}]{roldan2015}%
  \BibitemOpen
  \bibfield  {author} {\bibinfo {author} {\bibfnamefont {E.}~\bibnamefont
  {Roldan}}, \bibinfo {author} {\bibfnamefont {I.}~\bibnamefont {Neri}},
  \bibinfo {author} {\bibfnamefont {M.}~\bibnamefont {Dörpinghaus}}, \bibinfo
  {author} {\bibfnamefont {H.}~\bibnamefont {Meyr}},\ and\ \bibinfo {author}
  {\bibfnamefont {F.}~\bibnamefont {Jülicher}},\ }\bibfield  {title}
  {{\selectlanguage {en}\bibinfo {title} {Decision {Making} in the {Arrow} of
  {Time}}},\ }\href
  {https://doi.org/https://link.aps.org/doi/10.1103/PhysRevLett.115.250602}
  {\bibfield  {journal} {\bibinfo  {journal} {Physical Review Letters}\
  }\textbf {\bibinfo {volume} {115}},\ \bibinfo {pages} {250602} (\bibinfo
  {year} {2015})}\BibitemShut {NoStop}%
\bibitem [{\citenamefont {Neri}\ \emph {et~al.}(2017)\citenamefont {Neri},
  \citenamefont {Rold{\'a}n},\ and\ \citenamefont {J{\"u}licher}}]{neri17}%
  \BibitemOpen
  \bibfield  {author} {\bibinfo {author} {\bibfnamefont {I.}~\bibnamefont
  {Neri}}, \bibinfo {author} {\bibfnamefont {{\'E}.}~\bibnamefont
  {Rold{\'a}n}},\ and\ \bibinfo {author} {\bibfnamefont {F.}~\bibnamefont
  {J{\"u}licher}},\ }\bibfield  {title} {\bibinfo {title} {Statistics of infima
  and stopping times of entropy production and applications to active molecular
  processes},\ }\href {https://doi.org/10.1103/PhysRevX.7.011019} {\bibfield
  {journal} {\bibinfo  {journal} {Phys. Rev. X}\ }\textbf {\bibinfo {volume}
  {7}},\ \bibinfo {pages} {011019} (\bibinfo {year} {2017})}\BibitemShut
  {NoStop}%
\bibitem [{\citenamefont {Pigolotti}\ \emph {et~al.}(2017)\citenamefont
  {Pigolotti}, \citenamefont {Neri}, \citenamefont {Rold{\'a}n},\ and\
  \citenamefont {J{\"u}licher}}]{pigo17}%
  \BibitemOpen
  \bibfield  {author} {\bibinfo {author} {\bibfnamefont {S.}~\bibnamefont
  {Pigolotti}}, \bibinfo {author} {\bibfnamefont {I.}~\bibnamefont {Neri}},
  \bibinfo {author} {\bibfnamefont {{\'E}.}~\bibnamefont {Rold{\'a}n}},\ and\
  \bibinfo {author} {\bibfnamefont {F.}~\bibnamefont {J{\"u}licher}},\
  }\bibfield  {title} {\bibinfo {title} {Generic properties of stochastic
  entropy production},\ }\href {https://doi.org/10.1103/PhysRevLett.119.140604}
  {\bibfield  {journal} {\bibinfo  {journal} {Phys. Rev. Lett.}\ }\textbf
  {\bibinfo {volume} {119}},\ \bibinfo {pages} {140604} (\bibinfo {year}
  {2017})}\BibitemShut {NoStop}%
\bibitem [{\citenamefont {Seifert}(2019)}]{seif19}%
  \BibitemOpen
  \bibfield  {author} {\bibinfo {author} {\bibfnamefont {U.}~\bibnamefont
  {Seifert}},\ }\bibfield  {title} {\bibinfo {title} {From stochastic
  thermodynamics to thermodynamic inference},\ }\href
  {https://doi.org/10.1146/annurev-conmatphys-031218-013554} {\bibfield
  {journal} {\bibinfo  {journal} {Ann. Rev. Cond. Mat. Phys.}\ }\textbf
  {\bibinfo {volume} {10}},\ \bibinfo {pages} {171} (\bibinfo {year}
  {2019})}\BibitemShut {NoStop}%
\bibitem [{\citenamefont {Lucente}\ \emph {et~al.}(2022)\citenamefont
  {Lucente}, \citenamefont {Baldassarri}, \citenamefont {Puglisi},
  \citenamefont {Vulpiani},\ and\ \citenamefont {Viale}}]{lucente2022}%
  \BibitemOpen
  \bibfield  {author} {\bibinfo {author} {\bibfnamefont {D.}~\bibnamefont
  {Lucente}}, \bibinfo {author} {\bibfnamefont {A.}~\bibnamefont
  {Baldassarri}}, \bibinfo {author} {\bibfnamefont {A.}~\bibnamefont
  {Puglisi}}, \bibinfo {author} {\bibfnamefont {A.}~\bibnamefont {Vulpiani}},\
  and\ \bibinfo {author} {\bibfnamefont {M.}~\bibnamefont {Viale}},\ }\bibfield
   {title} {{\selectlanguage {en}\bibinfo {title} {Inference of time
  irreversibility from incomplete information: {Linear} systems and its
  pitfalls}},\ }\href {https://doi.org/10.1103/PhysRevResearch.4.043103}
  {\bibfield  {journal} {\bibinfo  {journal} {Physical Review Research}\
  }\textbf {\bibinfo {volume} {4}},\ \bibinfo {pages} {043103} (\bibinfo {year}
  {2022})}\BibitemShut {NoStop}%
\bibitem [{\citenamefont {Fodor}\ \emph {et~al.}(2016)\citenamefont {Fodor},
  \citenamefont {Nardini}, \citenamefont {Cates}, \citenamefont {Tailleur},
  \citenamefont {Visco},\ and\ \citenamefont {{van Wijland}}}]{fodo16}%
  \BibitemOpen
  \bibfield  {author} {\bibinfo {author} {\bibfnamefont {{\'E}.}~\bibnamefont
  {Fodor}}, \bibinfo {author} {\bibfnamefont {C.}~\bibnamefont {Nardini}},
  \bibinfo {author} {\bibfnamefont {M.~E.}\ \bibnamefont {Cates}}, \bibinfo
  {author} {\bibfnamefont {J.}~\bibnamefont {Tailleur}}, \bibinfo {author}
  {\bibfnamefont {P.}~\bibnamefont {Visco}},\ and\ \bibinfo {author}
  {\bibfnamefont {F.}~\bibnamefont {{van Wijland}}},\ }\bibfield  {title}
  {\bibinfo {title} {How far from equilibrium is active matter?},\ }\href
  {https://doi.org/10.1103/PhysRevLett.117.038103} {\bibfield  {journal}
  {\bibinfo  {journal} {Phys. Rev. Lett.}\ }\textbf {\bibinfo {volume} {117}},\
  \bibinfo {pages} {038103} (\bibinfo {year} {2016})}\BibitemShut {NoStop}%
\bibitem [{\citenamefont {Speck}(2016)}]{speck2016}%
  \BibitemOpen
  \bibfield  {author} {\bibinfo {author} {\bibfnamefont {T.}~\bibnamefont
  {Speck}},\ }\bibfield  {title} {{\selectlanguage {en}\bibinfo {title}
  {Stochastic thermodynamics for active matter}},\ }\href
  {https://doi.org/10.1209/0295-5075/114/30006} {\bibfield  {journal} {\bibinfo
   {journal} {Europhysics Letters}\ }\textbf {\bibinfo {volume} {114}},\
  \bibinfo {pages} {30006} (\bibinfo {year} {2016})}\BibitemShut {NoStop}%
\bibitem [{\citenamefont {Nardini}\ \emph {et~al.}(2017)\citenamefont
  {Nardini}, \citenamefont {Fodor}, \citenamefont {Tjhung}, \citenamefont {{van
  Wijland}}, \citenamefont {Tailleur},\ and\ \citenamefont {Cates}}]{nard17}%
  \BibitemOpen
  \bibfield  {author} {\bibinfo {author} {\bibfnamefont {C.}~\bibnamefont
  {Nardini}}, \bibinfo {author} {\bibfnamefont {{\'E}.}~\bibnamefont {Fodor}},
  \bibinfo {author} {\bibfnamefont {E.}~\bibnamefont {Tjhung}}, \bibinfo
  {author} {\bibfnamefont {F.}~\bibnamefont {{van Wijland}}}, \bibinfo {author}
  {\bibfnamefont {J.}~\bibnamefont {Tailleur}},\ and\ \bibinfo {author}
  {\bibfnamefont {M.~E.}\ \bibnamefont {Cates}},\ }\bibfield  {title} {\bibinfo
  {title} {Entropy production in field theories without time-reversal symmetry:
  {{Quantifying}} the non-equilibrium character of active matter},\ }\href
  {https://doi.org/10.1103/PhysRevX.7.021007} {\bibfield  {journal} {\bibinfo
  {journal} {Phys. Rev. X}\ }\textbf {\bibinfo {volume} {7}},\ \bibinfo {pages}
  {021007} (\bibinfo {year} {2017})}\BibitemShut {NoStop}%
\bibitem [{\citenamefont {Mandal}\ \emph {et~al.}(2017)\citenamefont {Mandal},
  \citenamefont {Klymko},\ and\ \citenamefont {DeWeese}}]{mand17}%
  \BibitemOpen
  \bibfield  {author} {\bibinfo {author} {\bibfnamefont {D.}~\bibnamefont
  {Mandal}}, \bibinfo {author} {\bibfnamefont {K.}~\bibnamefont {Klymko}},\
  and\ \bibinfo {author} {\bibfnamefont {M.~R.}\ \bibnamefont {DeWeese}},\
  }\bibfield  {title} {\bibinfo {title} {Entropy production and fluctuation
  theorems for active matter},\ }\href
  {https://doi.org/10.1103/PhysRevLett.119.258001} {\bibfield  {journal}
  {\bibinfo  {journal} {Phys. Rev. Lett.}\ }\textbf {\bibinfo {volume} {119}},\
  \bibinfo {pages} {258001} (\bibinfo {year} {2017})}\BibitemShut {NoStop}%
\bibitem [{\citenamefont {Pietzonka}\ and\ \citenamefont
  {Seifert}(2018)}]{piet17b}%
  \BibitemOpen
  \bibfield  {author} {\bibinfo {author} {\bibfnamefont {P.}~\bibnamefont
  {Pietzonka}}\ and\ \bibinfo {author} {\bibfnamefont {U.}~\bibnamefont
  {Seifert}},\ }\bibfield  {title} {\bibinfo {title} {Entropy production of
  active particles and for particles in active baths},\ }\href
  {https://doi.org/10.1088/1751-8121/aa91b9} {\bibfield  {journal} {\bibinfo
  {journal} {J. Phys. A: Math. Theor.}\ }\textbf {\bibinfo {volume} {51}},\
  \bibinfo {pages} {01LT01} (\bibinfo {year} {2018})}\BibitemShut {NoStop}%
\bibitem [{\citenamefont {Dabelow}\ \emph {et~al.}(2019)\citenamefont
  {Dabelow}, \citenamefont {Bo},\ and\ \citenamefont {Eichhorn}}]{dabe19}%
  \BibitemOpen
  \bibfield  {author} {\bibinfo {author} {\bibfnamefont {L.}~\bibnamefont
  {Dabelow}}, \bibinfo {author} {\bibfnamefont {S.}~\bibnamefont {Bo}},\ and\
  \bibinfo {author} {\bibfnamefont {R.}~\bibnamefont {Eichhorn}},\ }\bibfield
  {title} {\bibinfo {title} {Irreversibility in {{Active Matter Systems}}:
  {{Fluctuation Theorem}} and {{Mutual Information}}},\ }\href
  {https://doi.org/10.1103/PhysRevX.9.021009} {\bibfield  {journal} {\bibinfo
  {journal} {Phys. Rev. X}\ }\textbf {\bibinfo {volume} {9}},\ \bibinfo {pages}
  {021009} (\bibinfo {year} {2019})}\BibitemShut {NoStop}%
\bibitem [{\citenamefont {Flenner}\ and\ \citenamefont
  {Szamel}(2020)}]{flenner2020}%
  \BibitemOpen
  \bibfield  {author} {\bibinfo {author} {\bibfnamefont {E.}~\bibnamefont
  {Flenner}}\ and\ \bibinfo {author} {\bibfnamefont {G.}~\bibnamefont
  {Szamel}},\ }\bibfield  {title} {{\selectlanguage {en}\bibinfo {title}
  {Active matter: {Quantifying} the departure from equilibrium}},\ }\href
  {https://doi.org/10.1103/PhysRevE.102.022607} {\bibfield  {journal} {\bibinfo
   {journal} {Physical Review E}\ }\textbf {\bibinfo {volume} {102}},\ \bibinfo
  {pages} {022607} (\bibinfo {year} {2020})}\BibitemShut {NoStop}%
\bibitem [{\citenamefont {Gnesotto}\ \emph {et~al.}(2018)\citenamefont
  {Gnesotto}, \citenamefont {Mura}, \citenamefont {Gladrow},\ and\
  \citenamefont {Broedersz}}]{gnesotto2018}%
  \BibitemOpen
  \bibfield  {author} {\bibinfo {author} {\bibfnamefont {F.~S.}\ \bibnamefont
  {Gnesotto}}, \bibinfo {author} {\bibfnamefont {F.}~\bibnamefont {Mura}},
  \bibinfo {author} {\bibfnamefont {J.}~\bibnamefont {Gladrow}},\ and\ \bibinfo
  {author} {\bibfnamefont {C.~P.}\ \bibnamefont {Broedersz}},\ }\bibfield
  {title} {{\selectlanguage {en}\bibinfo {title} {Broken detailed balance and
  non-equilibrium dynamics in living systems: a review}},\ }\href
  {https://doi.org/10.1088/1361-6633/aab3ed} {\bibfield  {journal} {\bibinfo
  {journal} {Reports on Progress in Physics}\ }\textbf {\bibinfo {volume}
  {81}},\ \bibinfo {pages} {066601} (\bibinfo {year} {2018})}\BibitemShut
  {NoStop}%
\bibitem [{\citenamefont {Roldan}\ \emph {et~al.}(2021)\citenamefont {Roldan},
  \citenamefont {Barral}, \citenamefont {Martin}, \citenamefont {Parrondo},\
  and\ \citenamefont {Jülicher}}]{roldan2021}%
  \BibitemOpen
  \bibfield  {author} {\bibinfo {author} {\bibfnamefont {E.}~\bibnamefont
  {Roldan}}, \bibinfo {author} {\bibfnamefont {J.}~\bibnamefont {Barral}},
  \bibinfo {author} {\bibfnamefont {P.}~\bibnamefont {Martin}}, \bibinfo
  {author} {\bibfnamefont {J.~M.~R.}\ \bibnamefont {Parrondo}},\ and\ \bibinfo
  {author} {\bibfnamefont {F.}~\bibnamefont {Jülicher}},\ }\bibfield  {title}
  {{\selectlanguage {en}\bibinfo {title} {Quantifying entropy production in
  active fluctuations of the hair-cell bundle from time irreversibility and
  uncertainty relations}},\ }\href {https://doi.org/10.1088/1367-2630/ac0f18}
  {\bibfield  {journal} {\bibinfo  {journal} {New Journal of Physics}\ }\textbf
  {\bibinfo {volume} {23}},\ \bibinfo {pages} {083013} (\bibinfo {year}
  {2021})}\BibitemShut {NoStop}%
\bibitem [{\citenamefont {Barato}\ and\ \citenamefont
  {Seifert}(2015{\natexlab{b}})}]{bara15a}%
  \BibitemOpen
  \bibfield  {author} {\bibinfo {author} {\bibfnamefont {A.~C.}\ \bibnamefont
  {Barato}}\ and\ \bibinfo {author} {\bibfnamefont {U.}~\bibnamefont
  {Seifert}},\ }\bibfield  {title} {\bibinfo {title} {Universal bound on the
  {{Fano}} factor in enzyme kinetics},\ }\href
  {https://doi.org/10.1021/acs.jpcb.5b01918} {\bibfield  {journal} {\bibinfo
  {journal} {J. Phys. Chem. B}\ }\textbf {\bibinfo {volume} {119}},\ \bibinfo
  {pages} {6555} (\bibinfo {year} {2015}{\natexlab{b}})}\BibitemShut {NoStop}%
\bibitem [{\citenamefont {Pietzonka}\ \emph
  {et~al.}(2016{\natexlab{b}})\citenamefont {Pietzonka}, \citenamefont
  {Barato},\ and\ \citenamefont {Seifert}}]{piet16}%
  \BibitemOpen
  \bibfield  {author} {\bibinfo {author} {\bibfnamefont {P.}~\bibnamefont
  {Pietzonka}}, \bibinfo {author} {\bibfnamefont {A.~C.}\ \bibnamefont
  {Barato}},\ and\ \bibinfo {author} {\bibfnamefont {U.}~\bibnamefont
  {Seifert}},\ }\bibfield  {title} {\bibinfo {title} {Affinity- and
  topology-dependent bound on current fluctuations},\ }\href
  {https://doi.org/10.1088/1751-8113/49/34/34LT01} {\bibfield  {journal}
  {\bibinfo  {journal} {J. Phys. A: Math. Theor.}\ }\textbf {\bibinfo {volume}
  {49}},\ \bibinfo {pages} {34LT01} (\bibinfo {year}
  {2016}{\natexlab{b}})}\BibitemShut {NoStop}%
\bibitem [{\citenamefont {Satija}\ \emph {et~al.}(2020)\citenamefont {Satija},
  \citenamefont {Berezhkovskii},\ and\ \citenamefont {Makarov}}]{satija2020}%
  \BibitemOpen
  \bibfield  {author} {\bibinfo {author} {\bibfnamefont {R.}~\bibnamefont
  {Satija}}, \bibinfo {author} {\bibfnamefont {A.~M.}\ \bibnamefont
  {Berezhkovskii}},\ and\ \bibinfo {author} {\bibfnamefont {D.~E.}\
  \bibnamefont {Makarov}},\ }\bibfield  {title} {{\selectlanguage {en}\bibinfo
  {title} {Broad distributions of transition-path times are fingerprints of
  multidimensionality of the underlying free energy landscapes}},\ }\href
  {https://doi.org/10.1073/pnas.2008307117} {\bibfield  {journal} {\bibinfo
  {journal} {Proceedings of the National Academy of Sciences}\ }\textbf
  {\bibinfo {volume} {117}},\ \bibinfo {pages} {27116} (\bibinfo {year}
  {2020})}\BibitemShut {NoStop}%
\bibitem [{\citenamefont {Berezhkovskii}\ and\ \citenamefont
  {Makarov}(2021)}]{berezhkovskii2021}%
  \BibitemOpen
  \bibfield  {author} {\bibinfo {author} {\bibfnamefont {A.~M.}\ \bibnamefont
  {Berezhkovskii}}\ and\ \bibinfo {author} {\bibfnamefont {D.~E.}\ \bibnamefont
  {Makarov}},\ }\bibfield  {title} {{\selectlanguage {en}\bibinfo {title} {On
  distributions of barrier crossing times as observed in single-molecule
  studies of biomolecules}},\ }\href
  {https://doi.org/10.1016/j.bpr.2021.100029} {\bibfield  {journal} {\bibinfo
  {journal} {Biophysical Reports}\ }\textbf {\bibinfo {volume} {1}},\ \bibinfo
  {pages} {100029} (\bibinfo {year} {2021})}\BibitemShut {NoStop}%
\bibitem [{\citenamefont {Berezhkovskii}\ and\ \citenamefont
  {Makarov}(2019)}]{berezhkovskii2019}%
  \BibitemOpen
  \bibfield  {author} {\bibinfo {author} {\bibfnamefont {A.~M.}\ \bibnamefont
  {Berezhkovskii}}\ and\ \bibinfo {author} {\bibfnamefont {D.~E.}\ \bibnamefont
  {Makarov}},\ }\bibfield  {title} {\bibinfo {title} {On the forward/backward
  symmetry of transition path time distributions in nonequilibrium systems},\
  }\href {https://doi.org/10.1063/1.5109293} {\bibfield  {journal} {\bibinfo
  {journal} {The Journal of Chemical Physics}\ }\textbf {\bibinfo {volume}
  {151}},\ \bibinfo {pages} {065102} (\bibinfo {year} {2019})}\BibitemShut
  {NoStop}%
\bibitem [{\citenamefont {Berezhkovskii}\ \emph {et~al.}(2006)\citenamefont
  {Berezhkovskii}, \citenamefont {Hummer},\ and\ \citenamefont
  {Bezrukov}}]{berezhkovskii2006}%
  \BibitemOpen
  \bibfield  {author} {\bibinfo {author} {\bibfnamefont {A.}~\bibnamefont
  {Berezhkovskii}}, \bibinfo {author} {\bibfnamefont {G.}~\bibnamefont
  {Hummer}},\ and\ \bibinfo {author} {\bibfnamefont {S.}~\bibnamefont
  {Bezrukov}},\ }\bibfield  {title} {{\selectlanguage {en}\bibinfo {title}
  {Identity of {Distributions} of {Direct} {Uphill} and {Downhill}
  {Translocation} {Times} for {Particles} {Traversing} {Membrane}
  {Channels}}},\ }\href {https://doi.org/10.1103/PhysRevLett.97.020601}
  {\bibfield  {journal} {\bibinfo  {journal} {Physical Review Letters}\
  }\textbf {\bibinfo {volume} {97}},\ \bibinfo {pages} {020601} (\bibinfo
  {year} {2006})}\BibitemShut {NoStop}%
\end{thebibliography}%

\end{document}